\documentstyle[12pt,amssymb,amsmath]{article}
\begin{document}

{\Large

\noindent {\bf Unified $(q;\alpha,\beta,\gamma;\nu)$-deformation
of one-parametric}


\noindent{\bf  $q$-deformed oscillator algebras}}

\bigskip

\noindent{\bf I. M. Burban}

\noindent Bogolyubov Institute for Theoretical Physics, Kiev
03680, Ukraine

\noindent E-mail: burban@bitp.kiev.ua

\bigskip

\bigskip
\begin{abstract} We define a generalized
$(q;\alpha,\beta,\gamma;\nu)$-deformed oscillator algebra and study
the  number of its characteristics. We describe  the structure function of
deformation, analyze the classification of irreducible
representations and  discuss the asymptotic spectrum behaviour of the
Hamiltonian. For a special choice of the deformation parameters we
construct the deformed oscillator with discrete spectrum of its
"quantized coordinate" operator. We establish its connection with
the (generalized) discrete Hermite I polynomials.
\end{abstract}


\bigskip
\bigskip
\bigskip

Short title: Unified deformation
\bigskip\bigskip\bigskip

Corresponding author: Ivan Burban
\medskip

\noindent {\it Postal address}: Bogolyubov Institute for
Theoretical Physics, Metrologichna str. 14b,  Kiev 03680, Ukraine
 \bigskip

\noindent {\it Phone}: 380-44-5213436

\noindent {\it Fax}: 380-44-5265998

\newpage


\noindent
{\bf 1. Introduction}

\medskip

The investigation of an one-parameter deformed canonical
commutation relations in theoretical physics was originated from
the study of the dual resonance models of strong interactions. The
multi-parameter generalizations of these deformations were
actively studied with the appearance quantum groups, quantum
algebras, and quantum spaces. These deformations have found
applications in the study of exotic, different from the standard
Bose-Einstein and Fermi-Dirac, statistics.

On the other hand, many properties of the modified oscillator
algebra have found applications in the study of the integrability
of the two-particle Calogero models. The modified oscillator
algebra has been generalized to $C_{\lambda}$-extended oscillator
algebra \cite{Q} with the hope to exploit these deformations for
construction of new integrable models. With the same aim, a
"hybrid" model \cite{{V},{BHV}} of the modified and $q$-deformed
oscillator algebras has been proposed.

To close this cycle of ideas we consider the generalized
$(q;\alpha,\beta,\gamma;\nu)$-deformed oscillator algebra as "the
synthesis" of the $(q;\alpha,\beta,\gamma)$-deformed
\cite{{C-C-N-U},{B}} and $\nu$-modified oscillator algebras \cite
{BHV}.

In Section 2 some of the $q$-deformed oscillator algebras are
placed in the order of their extension. We find the structure
function of $(q;\alpha,\beta,\gamma;\nu)$-deformed oscillator
algebra. We show that this algebra is embedded into the deformed
$C_2$-extended oscillator algebra \cite{Q}. In Section 3 we give,
following \cite{Q, KMM}, classification of the representations of
this algebra. In Section 4 for the special choice of the
deformation parameters of the algebra we study the oscillator
\cite {Burb1} with the discrete spectrum of its "quantized
coordinate" $Q.$ This operator is represented by Jacobi matrix. We
find the eigenvalues and eigenfunctions of this operator. The
eigenfunctions are expressed by the (generalized) discrete
$q$-Hermite I polynomial \cite{Koe}. In Section 5 we represent the
Hamiltonian of this model as function of the number operator and
study its asymptotic energy spectrum behaviour.


\noindent

{\bf 2. Oscillator algebra and its generalized deformations}

\medskip

The oscillator algebra of the quantum harmonic oscillator is
defined by canonical commutation relations
\begin{equation}\label{burban: algeb} [a, a] = [a^+, a^+] = 0,\quad [a, a^+] = 1,\quad [N,
a] = - a,\quad[N, a^+]= a^+.
\end{equation} It allows the different types of deformations. Some of them
have been called {\it generalized deformed oscillator algebras}
\cite{{D1},{D2},{C-C-N-U},{MMP}}. Each of them defines an algebra
generated by elements (generators) $\{{\bf 1},a, a^+, N \}$ and
relations
\begin{equation}\label{burban: funct} [N, a]= - a,\quad [N, a^+] = a^+,
\quad a^+a = f(N), \quad aa^+ = f(N+1),
\end{equation} where $f$ is called  {\it a structure
function of the deformation}. Among them - the multiparameter
generalization of one-parameter deformations
\cite{{C-C-N-U},{MMP},{BDY},{BEM},{Q},{Bur},{MLD},{Burb1}}.

Let us recount some of them.

 1. The Arik-Coon $q$-deformed oscillator algebra \cite{AC}
 \[
 a a^+ - q a^+ a = 1,\quad [N,a]= - a,\quad [N, a^+] = a^+,\quad q\in{\mathbb R}_+,
 \]
 \begin{equation} f(n )= \frac{1-q^n}{1-q}.\end{equation}

2. The Biedengarn--Macfarlane $q$-deformed oscillator algebra
\cite{B,M}
\[ aa^+ - q a^+a = q^{- N},\quad aa^+ - q^{-1} a^+a = q^{N},\quad
 [N, a]= - a,\quad [N, a^+] = a^+,\]
 \begin{equation} f(n)=\frac{q^n - q^{-n}}{q - q^{-1}}\quad q\in{\mathbb
 R}_+.\end{equation}

3. The Chung-- Chung--Nam--Um generalized
$(q;\alpha,\beta)$-deformed oscillator algebra \cite{C-C-N-U}
 \[ aa^+ - q a^+a = q^{\alpha N + \beta},\quad
[N, a]= - a,\quad [N, a^+] = a^+,\quad q\in{\mathbb R}_+,\quad
\alpha, \beta \in{\mathbb R},\]\begin{equation} f(n)=
\begin{cases}
q^{\beta}\frac{q^{\alpha n} - q}{q^{\alpha} - q}, &\text {if
$\alpha \ne 1$}\\ nq^{n-1 + \beta}, &\text {if $\alpha = 1.$}\\
\end{cases}
\end{equation}

4. The generalized $(q;\alpha,\beta,\gamma)$-deformed oscillator
algebra \cite{BDY} \[ aa^+ - q^{\gamma} a^+a = q^{\alpha N +
\beta},\quad [N, a]= - a,\quad [N, a^+] = a^+,\quad q\in{\mathbb
R}_+,\alpha, \beta,\gamma \in{\mathbb R},\]
\begin{equation}\label{burban: bor}
f(n)=
\begin{cases}
q^{\beta}\frac{q^{\alpha n}- q^{\gamma n}}{q^{\alpha}- q^{\gamma
n}}, &\text {if $\alpha \ne \gamma $}\\ nq^{n-1 + \beta}, &\text
{if $\alpha = \gamma.$}\\
\end{cases}
\end{equation}

5. The $\nu$-modified oscillator algebra \cite {{V},{BHV}}
\[ [a, a^+]= 1 + 2\nu K,\quad [N, a]= - a,\quad [N, a^+] =
a^+,\]\[\quad a K = - K a,\quad a^+ K = - K a^+,\quad K^2 =
1,\quad \nu \in \mathbb R,\]
\begin{equation}\label{burban: vas}
f(n)=
\begin{cases}
2k + 1 + 2\nu, &\text {if $ n = 2k $}\\ 2k+2, &\text {if $n =
2k+1$}.\\
\end{cases}
\end{equation}
This oscillator, as it has been shown in \cite{BHV}, is connected
with the two-particle Calogero model \cite {Cal}.

6. The deformed $C_{\lambda}$-extended oscillator algebra \cite
{Q} is defined by the relations \[ [a, a^+]_q \equiv aa^+ - q a^+a
= H(N) + K(N)\sum _{k = 0}^{\lambda -1}\nu_k P_k ,\quad [N, a]= -
a,\quad [N, a^+] = a^+,\]\begin{equation}\label{burban: Ques}
\quad a K = - K a,\quad a^+ K = - K a^+,\quad K^2 = 1,\quad \nu_k
\in \mathbb R,\end{equation} where  $\nu_k\in \mathbb R$ and
$H(K),\, K(N)$ are real analytic functions. This algebra admits
the two Casimir operators $C_1= e^{2\pi N} and \quad C_2 = \sum_{k
= 0}^{\lambda -1}e^{-2\pi i(N-k)/\lambda}P_k.$

7. The new $(q;\nu)$-deformed oscillator \cite {BEM} \[ aa^+ - q
a^+ a = (1 + 2\nu K) q^{-N},\quad [N, a]= - a,\]\[ [N, a^+] =
a^+,\quad K a = - a K,\quad K a^+ = - a^+ K,\quad K^2 =
1,\]\begin{equation} f(n)= \Bigl(\frac{q^n - q^{-n}}{q - q^{-1}}+
2\nu \frac {q^n - (-1)^n q^{-n}}{q + q^{-1}}\Bigr)\end{equation}
has been defined by the combination of the idea of Biedenharn--
Macfarlane \cite {{B},{M}} $q$-deformation with Brink, Hanson and
Vasiliev idea \cite {BHV} of the $\nu$-modification of the
oscillator algebra.

8. In order to close this cycle of ideas we consider a
$(q;\alpha,\beta,\gamma;\nu)$-deformed oscillator algebra --
("hybrid" of the $(q;\alpha,\beta,\gamma)$-deformed (\ref {burban:
bor}) and the $\nu$-modified (\ref{burban: vas}) oscillator
algebras ), or, more exactly, an oscillator defined by generators
$\{I, a, a^+, N, K\}$ and relations
\[ aa^+ - q^{\gamma} a^+a =(1 + 2\nu K) q^{\alpha N +
\beta},\quad [N, a]= - a,\quad [N, a^+] = a^+,
\]
\begin{equation}\label{burban: first}  K a = - a ,\quad K a^+ = - a^+ K,\quad
[N, K] = 0,\quad N^+ = N,\quad K^+ = K,
\end{equation}
where $q \in {\mathbb R}_+,\alpha,\beta \in {\mathbb R}, \nu \in
{\mathbb R}-\{0\}.$ This model unify all deformations 1. - 7. of
the oscillator algebra (\ref{burban: algeb}).


\noindent

{\bf 3. Generalized $(q;\alpha,\beta,\gamma;\nu)$-deformed
oscillator algebra and its simplest properties}

\medskip

{\it (a) $(q;\alpha,\beta,\gamma;\nu)$-deformed structure
function.} Description of an deformed oscillator algebra requires
the determination of the deformation structure function $f(n).$

Equations (\ref{burban: funct}) and (\ref{burban: first}) imply
the recurrence relation
\begin{equation}\label{burban: struct0} f(n+1) - q^{\gamma} f(n) =
\Bigl(1 + 2 \nu(-1)^n\Bigr)q^{\alpha n + \beta}. \end{equation}
Its solution is obtained by method of mathematical induction:

\noindent $0.\quad f(1) = q^{\gamma} f(0) + \Bigr(1 + 2
\nu(-1)^0\Bigr)q^{\alpha 0 + \beta};$
\[1.\quad f(2) = q^{\gamma} f(1) + \Bigr(1 - 2 \nu(-1)^1\Bigr)q^{\alpha 1
+ \beta} = [q^{2\gamma} f(0) + \Bigr(1 + 2
\nu(-1)^0\Bigr)q^{\gamma}q^{\alpha 0 + \beta} + \Bigr(1 - 2
\nu(-1)^1\Bigr)q^{\alpha 1 + \beta};\]
\[2.\quad f(3) = q^{\gamma} f(2) + \Bigr(1 - 2 \nu(-1)^2\Bigr)q^{\alpha
2 + \beta} = q^{3\gamma} f(0) + \Bigr(1 + 2
\nu(-1)^0\Bigr)q^{2\gamma}q^{\alpha 0 + \beta} + \Bigr(1 - 2
\nu(-1)^1\Bigr)q^{\gamma}q^{\alpha 1 + \beta} + \]$$\Bigr(1 - 2
\nu(-1)^2\Bigr)q^{\alpha 2 + \beta}; $$
\[3.\quad f(4) = q^{\gamma} f(3) + \Bigr(1 + 2 \nu(-1)^3\Bigr)q^{\alpha 3
+ \beta} = q^{4\gamma} f(0) + \Bigr(1 + 2
\nu(-1)^0\Bigr)q^{3\gamma}q^{\alpha 0 + \beta} + \Bigr(1 - 2
\nu(-1)^1\Bigr)q^{2\gamma}q^{\alpha 1 + \beta} + \] $\Bigr(1 - 2
\nu(-1)^2\Bigr)q^{\gamma}q^{\alpha 2 + \beta} + \Bigr(1 + 2
\nu(-1)^3\Bigr)q^{\alpha 3 + \beta};$

\noindent $.$

\noindent$.$

\noindent $.$
\[n.\quad f(n) = q^{n\gamma} f(0) + \sum_{k = 0}^{n-1}q^{\gamma(n-k-1)}q^{\alpha k + \beta} +
2 \nu \sum_{k=0}^{n-1}q^{\gamma(n-k-1)}(-1)^{k}q^{\alpha
k+\beta}.\qquad \qquad {}{}{}\] The solution of the equation
(\ref{burban: struct0}) with the initial value $f(0)$ is given by
formula
\begin{equation} \label{burban: struct} f(n) =
\begin{cases} f(0)q^{\gamma n} +
q^{\beta}\Bigl(\frac {q^{\gamma n} - q^{\alpha n}} {q^{\gamma} -
q^{\alpha}} + 2\nu \frac {q^{\gamma n} - (-1)^n q^{\alpha n}}
{q^{\gamma} + q^{\alpha}}\Bigr),&\text{if $\alpha \ne \gamma$}\\
 f(0)q^{\gamma n} + n q^{\gamma(n-1) + \beta} + 2
\nu q^{\gamma(n-1) + \beta} \Bigl(\frac{1 -
(-1)^n}{2}\Bigr),&\text{ if $\alpha = \gamma.$}\end{cases}
\end{equation}

{\it (b) Positivity of $(q;\alpha,\beta,\gamma;\nu)$-deformed
structure function.} We try to define the values of the parameters
$\{\alpha,\beta, \gamma,\nu \},$ where function $ f(n)$ is
positive.

The inequality
\begin{equation}\label{burban: ine}\frac{q^{\gamma n} - q^{\alpha n}}{q^{\gamma} -
q^{\alpha}} + 2\nu \frac{q^{\gamma n} - (-1)^n q^{\alpha n}}
{q^{\gamma} + q^{\alpha}} > 0,\end{equation} can be rewritten as
\begin{equation}
(q^{\gamma n}- q^{\alpha n})\Bigl(\frac{1}{q^{\gamma} -
q^{\alpha}} + 2 \nu\frac{1}{q^{\gamma} + q^{\alpha}}\Bigr) > 0
\end{equation}
for even $n$ and
\begin{equation}\frac{q^{\gamma n}- q^{\alpha n}}{q^{\gamma n}+ q^{\alpha
n}} \frac{q^{\gamma }+ q^{\alpha}}{q^{\gamma }- q^{\alpha }} + 2
\nu
> 0
\end{equation}
for odd $n.$

It follows \begin{equation}\label{burban: fre}1 + 2 \nu > 0
\end{equation} if $q
> 1, \gamma - \alpha > 0 \,( q < 1,\, \gamma - \alpha < 0)$

\begin{equation}\label{burban: frest} -1 < 2 \nu < -\frac{q^{\gamma} +
q^{\alpha}}{q^{\gamma} - q^{\alpha}}
\end{equation}
if $q < 1, \gamma - \alpha
> 0 \,( q > 1,\, \gamma - \alpha < 0).$

The conditions (\ref{burban: fre}) and (\ref{burban: frest}) are
sufficient to ensure that the Fock representation of the relation
(\ref{burban: first}) has the Hermitian properties.

{\it (c) Useful formulas.} The two formulae for the study of this
algebra will be useful. One of them
\begin{equation}\label{burban: useful} a(a^+)^n - q^{\gamma n }\, (a^+)^n a =
[n;\alpha,\gamma;\nu K](a^+)^{n-1}q^{\alpha
N+\beta},\end{equation} where $n\ge 1,$
\begin{equation}  [n;\alpha,\gamma;\nu K] =
\begin{cases}
\Bigl(\frac {q^{\gamma n} - q^{\alpha n}} {q^{\gamma} -
q^{\alpha}}+ 2\nu K\frac {q^{\gamma n} - (-1)^n q^{\alpha n}}
{q^{\gamma} + q^{\alpha}}\Bigr),&\text{if $\alpha \ne \gamma$;}\\
 n q^{\alpha(n-1)} + 2\nu K q^{\alpha(n-1)} \Bigl(\frac{1 - (-1)^n}{2}\Bigr),&\text{
if $\alpha = \gamma$}\end{cases}
\end{equation}
is deduced by method of mathematical induction.

 Indeed, for $n = 1$ the relation (\ref{burban: useful}) is true by
definition. The assumption that it is true for some $n$ implies
\[ \{ a(a^+)^{n+1} = a(a^+)^n a^+ =
\Bigl( q^{\gamma n}(a^+)^n a+[n;\alpha,\gamma;\nu K
](a^+)^{n-1}q^{\alpha N + \beta}\Bigr) \, a^+ = \]
\[ q^{\gamma n}(a^+)^n a a^+ + (a^+)^{n}[n;\alpha,\gamma;-\nu K]q^{\alpha
(N+1)+ \beta} = \] \[q^{\gamma n}(a^+)^n \Bigl(q^{\gamma} a^+ a +
(1 + 2\nu K )q^{\alpha N + \beta}\Bigr) + (a^+)^{n}
[n;\alpha,\gamma;-\nu K] q^{\alpha (N+1) + \beta} = \] \[
q^{\gamma (n+1)}(a^+)^{(n+1)}a + q^{\gamma n}(a^+)^n \Bigl(1 + 2
\nu K \Bigr) q^{\alpha N + \beta} + q^{\alpha}(a^+)^n
[n;\alpha,\gamma;-\nu K] q^{\alpha N + \beta}.\] The direct
calculations leads to (\ref{burban: useful}).

The second formula gives the generated function for
$[n;\alpha,\gamma;\nu K]:
$
\begin{equation}
\sum_{n = 0}^{\infty}[n;\alpha,\gamma;\nu K]z^n =
\begin{cases}
\frac{z}{1-q^{\gamma}z}\Bigl(\frac{1}{1-q^{\alpha}z}+2\nu K
\frac{1}{1+q^{\alpha}z}\Bigr) , &\text {if $\alpha \ne \gamma$}\\
\frac{z}{(1-q^{\gamma}z)^2}+ 2\nu K\frac{z}{1-q^{2\gamma}z^2} ,
&\text {if $\alpha = \gamma.$}\\
\end{cases}
\end{equation}

{\it (d) Deformed $C_2$-extended and
$(q;\alpha,\beta,\gamma;\nu)$-deformed oscillator algebras.}
 The defining relations of the deformed $C_2$-extended oscillator
are given by $$ [N, a^+]= a^+,\quad [N,P_{k}] = 0,\quad a^+ P_{k}
= P_{k +1}a^+,\quad P_1 + P_2 = I,\quad P_{k} P_{l} =
\delta_{k,l}P_{l},$$
\begin{equation}\label{burban: deform}
aa^+ - q^{\gamma} a^+ a = H(N)+ \nu\Bigl( E(N+1) + q^{\gamma}
E(N)\Bigr)( P_0 - P_1),
\end{equation}
where $q, \nu \in {\mathbb R},k,l =1, 2,$ and $E(N), H(N)$ are
real analytic functions. As we saw above the deformed extended
oscillator algebra $C_{\lambda}$ admits the two Casimir operators
$C_1, C_2.$ In case of the $C_2$-extended oscillator algebra they
have the form
\begin{equation} C_1 = e^{2\pi i N},\quad C_2 =
\Sigma_{k = 0}^1e^{-2\pi i(N-k)/2}= e^{i\pi N}K.
\end{equation}
Let us define the operator \begin{equation}\label{burban: casim}
{\tilde C}_3 = q^{-\gamma N}\Bigl(D(N) + \nu E(N) K -
a^+a\Bigr),\end{equation} where $D(N),\, E(N)$ some analytic
functions of $N$. The operator ${\tilde C}_3$ will be the Casimir
operator of the oscillator algebra (\ref{burban: deform}) if the
only one condition $[{\tilde C}_3, a] = 0$ holds.  It amounts to
determination of the solution of the equations
\begin{equation}\label{burban: addi}
D(N+1)- q^{\gamma}D(N) = H(N),\quad E(N+1)\beta_{k+1}-
q^{\gamma}E(N)\beta_{k} = K(N)\nu_{k},
\end{equation} where $\nu_0 = - \nu_1 =\nu, \beta_0 = 0,\beta_2 =
0, \beta_1 = \nu,\quad k = 0, 1.$ Putting the solution $ E(N) = 2
q^{\alpha N + \beta}/({q^{\gamma} + q^{\alpha}})$ of the equation
of (\ref{burban: addi}) and $ H(N) = q^{\alpha N + \beta}$ in
(\ref{burban: deform}), we obtain the commutation relations of the
$ (q;\alpha,\beta,\gamma;\nu)$-deformed oscillator algebra
(\ref{burban: first}). Moreover, the solution $$ D(N) =
\begin{cases} q^{\beta}\Bigl(\frac{q^{\gamma N } - q^{\alpha N
}}{q^{\gamma} - q^{\alpha}} + 2\nu\frac{q^{\gamma N}}{q^{\gamma} +
q^{\alpha}}\Bigr)&\text{if $\gamma \ne \alpha
$}\\q^{\beta}(q^{\gamma(N-1)}N + \nu q^{-\gamma})&\text{if $\gamma
=\alpha $}\end{cases} $$ of the first equation (\ref{burban:
addi}) gives the explicit form of the Casimir operator
\begin{equation}\label{burban: casimir}
{\tilde C}_3 = \begin {cases}
q^{-\gamma N }\Bigl((\frac{q^{\gamma N} - q^{\alpha N
}}{q^{\gamma} - q^{\alpha}} + 2 \nu \frac{q^{\gamma N} - (-1)^N
q^{\alpha N}}{q^{\gamma} + q^{\alpha}})q^{\beta} - a^+a
\Bigr)&\text {if $\alpha\ne\gamma$}\\ q^{-\gamma N}\Bigl(N + \nu(1
+ (-1)^N)q^{\gamma N + \beta}- a^+a\Bigr)&\text{if $\alpha =
\gamma $}.\end{cases}\end{equation}

\noindent

{\bf 3. Classification of representations of unified
$(q;\alpha,\beta,\gamma;\nu)$-deformed oscillator algebra}

\medskip

As has been shown in the previous Section the
$(q;\alpha,\beta,\gamma;\nu)$-deformed oscillator algebra allows
nontrivial center which means that it has irreducible
non-equivalent representations \cite {{Rid},{Ch}}. We give a
classification of these representations by method similar to the
one in the articles \cite{Q}, \cite {KMM}.

Due to the relations (\ref{burban: first}) there exists a vector
$|0\rangle$ such that $$a^+a|0\rangle = \lambda_0 |0\rangle, \quad
aa^+ |0\rangle = \mu_0|0\rangle,\quad N |0\rangle = \varkappa_0
|0\rangle,\quad K|0\rangle = \omega e^{-i\pi
\varkappa_0}|0\rangle, $$ where $\omega$ is the value of the
Casimir operator $C_2$ on the given irreducible representation. By
using the formula (\ref{burban: useful}) we find that vectors
\begin{equation}
|n\rangle = \begin{cases} (a^+)^n|0\rangle, &\text{if $n\ge 0$}\\
(a)^{-n}|0\rangle, &\text{if $n < 0$}\\
\end{cases}
\end{equation}
are eigenvectors of the operators $a^+a$ and $aa^+:$
\[ a^+ a|n\rangle = \lambda_n |n\rangle, \quad aa^+ |n\rangle = \mu_n |n\rangle.\]

Let us define new system of the orthonormal vectors
$\{|n\rangle\}_{n = - \infty}^{n = \infty},$ by
\begin{equation}
|n\rangle = \begin{cases}\Bigl(\prod_{k = 1}^{n} \lambda_k
\Bigr)^{-1/2} (a^+)^n|0\rangle, &\text{if $n\ge 0$}\\
\Bigl(\prod_{k = 1}^{-n}\lambda_{n+k}\Bigr)^{-1/2}
(a)^{-n}|0\rangle, &\text{if $n < 0$}.\\
\end{cases}
\end{equation}
Then the relations (\ref{burban: first}) are represented by
operators
\[
a^+|n\rangle = \sqrt{\lambda_{n+1}}|n+1\rangle, \quad a |n\rangle
= \sqrt{\lambda_n}|n-1\rangle, \]
\begin{equation}\label{burban: rep} N|n\rangle = (\varkappa_0 +
n)|n\rangle, \quad K |n\rangle =
\frac{(-1)^n}{2\nu}B|n\rangle,\end{equation} where $B = 2 \nu
\omega e^{-i\pi \varkappa_0}\in {\mathbb R}.$ Due to
non-negativity of the operators $a^+a,$ $aa^+$ we have
$\lambda_n\ge 0$ and $\mu_n\ge 0.$ By using (\ref{burban: first}),
we find (\ref{burban: rep}) and $\lambda_n =\mu_{n-1}$ we obtain
the recurrence relation
\begin{equation}\label{burban: resur}
\lambda_{n+1} - q^{\gamma}\lambda_n = \Bigl(1 + (-1)^n B
\Bigr)q^{\alpha(n +\varkappa_0)+ \beta}.
\end{equation}
Take into account the relation (\ref{burban: struct}) the solution
of equation (\ref{burban: resur}) can be represented by
\begin{equation} \label{burban: solu} \lambda_n =
\begin{cases} \lambda_0 q^{\gamma n} +
q^{\alpha\varkappa_0 + \beta}\Bigl(\frac {q^{\gamma n} - q^{\alpha
n}} {q^{\gamma} - q^{\alpha}} + B \frac {q^{\gamma n} - (-1)^n
q^{\alpha n}} {q^{\gamma} + q^{\alpha}}\Bigr),&\text{if $\alpha
\ne \gamma$;}\\ \lambda_0 q^{\gamma n} + n q^{\gamma(n +
\varkappa_0 -1) + \beta } + B q^{\gamma(n + \varkappa_0 -1)+
\beta} \Bigl(\frac{1-(-1)^n}{2}\Bigr),&\text{ if $\alpha =
\gamma.$}\end{cases}\end{equation}

Equivalently, using the expression of the Casimir operator
(\ref{burban: casimir}) and the representation (\ref{burban: rep})
the solution $\lambda_n$ of equation (\ref{burban: resur}) can be
presented by means of the eigenvalue $c_3$ of the Casimir operator
$C_3$ in irreducible representation of the relation (\ref{burban:
first})
\begin{equation}\label{burban: coins}
\lambda_n =
\begin{cases}  - q^{\gamma n}c_3 +
q^{\beta}\Bigl(\frac {q^{\gamma (n + \varkappa_0)} - q^{\alpha( n
+ \varkappa_0)}} {q^{\gamma} - q^{\alpha}} + 2\nu \frac
{q^{\gamma(n +\varkappa_0) } - (-1)^{n + \varkappa_0} q^{\alpha
(n+\varkappa_0)}} {q^{\gamma} + q^{\alpha}}\Bigr),&\text{if
$\alpha \ne \gamma$;}\\ \lambda_0 q^{\gamma n}c_3 + n q^{\gamma(n
+ \varkappa_0 -1) + \beta } + B q^{\gamma(n + \varkappa_0 -1)+
\beta} \Bigl(\frac{1-(-1)^n}{2}\Bigr),&\text{ if $\alpha =
\gamma.$}\end{cases}\end{equation} It is easy see that $\lambda_n$
in (\ref{burban: solu}) coincides with the value of (\ref{burban:
coins}), where $B = (-1)^{\varkappa_{0}}2\nu.$

According to \cite {Q} a representations of the generalized
oscillator algebra is reduced to the four classes of unireps:

1. Representations {\it bounded from below}. They are defined by
(\ref{burban: rep}) and $n_1\in {\mathbb Z}_{\le 0}$ such that
\begin{equation}\label{burban: clas1} \lambda_n =
\begin{cases}
\lambda_{n_1} = 0 &\text {if $n_1\in \{\ldots, -2, -1, 0
\}$}\\\lambda_n > 0 &\text {if $n \in \{n_1+1, n_1+2,
\ldots\}$}.\\
\end{cases}
\end{equation}

(i) Let $\lambda_n$ be defined by formula (\ref{burban: solu}) for
$\gamma - \alpha = 0,$ $q > 0$. The non-negativity of $\lambda_n$
implies
\begin{equation}\label{burban: sabst} \lambda_n = \lambda_0 + n
q^{\gamma( \varkappa_0 -1) + \beta} + B q^{\gamma( \varkappa_0 -1)
+ \beta} \Bigl(\frac{1 - (-1)^n}{2}\Bigr)\ge 0.
\end{equation}
It follows (\ref{burban: clas1}) and due to (\ref{burban: rep}) $a
|n_1\rangle = 0.$

After possible renumbering of vectors $|n\rangle$ we obtain
$a|0\rangle = 0$ and due to (\ref{burban: rep}) one gets
$\lambda_0 = 0.$ Therefore the representations are given by
formulae (\ref{burban: rep}) with
\begin{equation}\lambda_n = n q^{\gamma(n + \varkappa_0 -1) +
\beta} + B q^{\gamma(n + \varkappa_0 -1) + \beta}\Bigl(\frac{1 -
(-1)^n}{2}\Bigr),\end{equation} where $ n \ge 0$. The arbitrary
values of the parameter $\varkappa_0,$ and $B\ge 0$ correspond to
nonequivalent representations of (\ref{burban: first}).

(ii) Let $\lambda_n$ be defined by (\ref{burban: solu}) for
$\gamma - \alpha >0,$ $q > 1$ $(\gamma-\alpha < 0, 0 < q < 1).$

The assumption of the positivity at last one of the numbers
$\frac{1}{q^{\gamma} - q^{\alpha}} \pm \frac{ B}{q^{\gamma} +
q^{\alpha}}$ and the nonnegativity of  $\lambda_n$
\begin{equation}\label{burban: cs1}
\Bigl(\lambda_0 q^{-(\alpha \varkappa + \beta)}+
\frac{1}{q^{\gamma} - q^{\alpha}} + \frac{B}{q^{\gamma} +
q^{\alpha}}\Bigr)\ge q^{-(\gamma
-\alpha)n}\Bigl(\frac{1}{q^{\gamma} - q^{\alpha}} + \frac{(-1)^n
B}{q^{\gamma} + q^{\alpha}} \Bigr)
\end{equation} for $n\ge 0$
implies (\ref{burban: clas1}) and and due to (\ref{burban: rep})
$a |n_1\rangle = 0.$

After possible renumbering of vectors $|n\rangle$ we obtain
$a|0\rangle = 0$ and due to (\ref{burban: rep}) one gets
$\lambda_0 = 0.$ Therefore the representations are given by
(\ref{burban: rep}) with
\begin{equation}\label{burban: so1}
\lambda_n = q^{\alpha\varkappa_0  + \beta + n\gamma}\Bigl(\frac{1
- q^{(\alpha -\gamma)n}}{q^{\gamma} - q^{\alpha}}+ B\frac{1
-(-1)^nq^{(\alpha - \gamma)n}}{q^{\gamma} +
q^{\alpha}}\Bigr),\quad n \ge 0.
\end{equation}
The nonnegativity condition for $\lambda_n$ implies $B \ge -1.$

The arbitrary values of the parameter $\varkappa_0,$ and
$\lambda_0 = 0$ and $B > -1$ defined nonequivalent
infinite-dimensional representations of (\ref{burban: first}).

iii) Let $\lambda_n$ be given by (\ref{burban: solu}) for $\gamma
- \alpha > 0,$ $0 < q < 1 (\gamma-\alpha < 0, q > 1).$

It can be represented by $$\lambda_n = q^{\alpha\varkappa_0 +
\beta + \gamma n}\times $$
\begin{equation}\label{burban: fin}
\Bigl\{\Bigl(\lambda_0 q^{-(\alpha \varkappa + \beta)}+
\frac{1}{q^{\gamma} - q^{\alpha}} + \frac{B}{q^{\gamma} +
q^{\alpha}}\Bigr) - q^{-(\gamma
-\alpha)n}\Bigl(\frac{1}{q^{\gamma} - q^{\alpha}} + \frac{(-1)^n
B}{q^{\gamma} + q^{\alpha}} \Bigr)\Bigr\}.\end{equation}

The nonnegativity of $\lambda_n\ge 0,$ negativity of
$\Bigl(\lambda_0 q^{-(\alpha \varkappa + \beta)} +
\frac{1}{q^{\gamma} - q^{\alpha}} + \frac{B}{q^{\gamma} +
q^{\alpha}}\Bigr)< 0,$ and the non-positivity of
$\Bigl(\frac{1}{q^{\gamma} - q^{\alpha}} + \frac{(-1)^n
B}{q^{\gamma} + q^{\alpha}} \Bigr)$ implies (\ref{burban: clas1})
and $a |n_1\rangle = 0.$ The same arguments as in the items i),
ii) give $a|0\rangle = 0$ and due to (\ref{burban: rep})
$\lambda_0 = 0.$ Therefore the representation is given by (\ref
{burban: rep}) with $\lambda_n$ as (\ref{burban: so1}). The
nonnegativity condition for $\lambda_n$ gives a restriction for
possible values of $B:$
\begin{equation}- 1 \le B < -\frac{q^{\gamma} + q^{\alpha}}{q^{\gamma} +
q^{\alpha}}
\end{equation}
 The arbitrary values of the
parameter $\varkappa_0,$ and $ -1\le B < - \frac{q^{\gamma}+
q^{\alpha}}{q^{\gamma} + q^{\alpha}}$ and $\lambda_0 = 0 $
distinguished irreducible representation of the relations
(\ref{burban: first}).

 2. Representations {\it bounded from above.}
They are defined by formulae (\ref{burban: rep}) and $n_2\in
{\mathbb N}$ such that
\begin{equation} \label{burban: clas2} \lambda_n =
\begin{cases}
\lambda_{n_2} = 0 &\text {if $n_2\in \{1, 2, 3, \ldots \}$}
\\\lambda_n
> 0 &\text {if $n \in \{n_2-1, n_2-2, \ldots\} $}.\\
\end{cases}
\end{equation}
Let $\lambda_n$ be given by formula (\ref{burban: solu}) $0 < q <
1, (\alpha - \gamma) > 0 \quad (q > 1, (\alpha - \gamma) < 0 $)
and both values $\frac{1}{q^{\gamma} - q^{\alpha}} \pm
\frac{B}{q^{\gamma} + q^{\alpha}}$ are non-positive (at last one
of them must be strictly negative). From the non-negativity
condition $\lambda_n\ge 0$ it follows the existence of $n_2$ such
that (\ref{burban: clas2}) holds. From formulae (\ref{burban:
rep}) we obtain $a^+ |n_2\rangle = 0,$ and after possible
renumbering we may assume that $a^+ |0\rangle = 0,$ and due to
(\ref{burban: rep}) $\lambda_1 = 0.$

The formula (\ref{burban: sabst}) implies $\lambda_0 = -
q^{\alpha\varkappa_0 + \beta - \gamma}(1+B).$  The condition
$\lambda_0\ge 0$ is equivalent to $B \le -1.$ It follows
\begin{equation}\label{burban: solu2}
\lambda_n =  q^{\alpha \varkappa_0 + \beta + \gamma
n}\Bigl(\Bigl(\frac{1- q^{(\alpha - \gamma)n}}{q^{\gamma} -
q^{\alpha}} - q^{-\gamma}\Bigr) + B \Bigl(\frac{1- q^{(\alpha
-\gamma)n}(-1)^n}{q^{\gamma} + q^{\alpha}} - q^{-\gamma} \Bigr)\ge
0,
\end{equation}
$n \le 0.$ Therefore the representation is given by formulae (\ref
{burban: rep}) with $\lambda_n$ as (\ref{burban: solu2}). If $B <
-1$ the nonnegativity condition  for $\lambda_n$ gives a
restriction for possible values of $B:$
\begin{equation} B \le -\frac{q^{\gamma}+ q^{\alpha}}{q^{\gamma}+
q^{\alpha}}
\end{equation}
 The arbitrary values of parameter $\varkappa_0$ and $\lambda_0 = -
q^{\alpha\varkappa_0 + \beta - \gamma}(1+B),$ $B <
\frac{q^{\gamma}+ q^{\alpha}}{q^{\gamma}+ q^{\alpha}} $
distinguish irreducible representation of (\ref{burban: first}).

3. The {\it finite dimensional representation.} They defined by
formulae (\ref{burban: rep}) and $n_1\in\{\ldots,-2,-1,0\},$
$n_2\in \{ 1,2,3,\ldots\}$ such that
\begin{equation}\label{burban: findim1} \lambda_n =
\begin{cases}
\lambda_{n_1} = \lambda_{n_2} = 0 &\text {if $n_2\in \{1, 2, 3,
\ldots \}, n_1\in \{ \ldots, -2, -1, 0\}$}
\\ \lambda_n > 0 &\text {if $n \in\{n_1 +1, n_1+2, \ldots, n_2-1\}$}.\\
\end{cases}
\end{equation}

i) One-dimensional representations. These representations are
given by formulae (\ref{burban: rep}) with $\lambda_n$ calculated
from (\ref{burban: so1}), (\ref{burban: fin}) and (\ref{burban:
solu2}) at $B = -1$
\begin{equation}
\lambda_n = q^{-\alpha \varkappa_0 + \beta }\Bigl(\frac{q^{\gamma
n }-q^{\alpha n}}{q^{\gamma}-q^{\alpha}} - \frac{q^{\gamma n} -
(-1)^n q^{\alpha n}}{q^{\gamma} + q^{\alpha}}\Bigr).
\end{equation}
In this case $n_1 = 0$ $(n_1 = -1)$ and $n_2 = 1$ $(n_2 = 0)$ and
the representations (\ref{burban: rep}) are defined by
\begin{equation}\label{burban: onedim}
a = a^+ = 0, \quad N = \varkappa_0,\quad K = - \frac{1}{2\nu}.
\end{equation}
These representations are one-dimensional. Their are parametrized
by $\varkappa_0.$

ii) Two-dimensional representations. Let $\lambda_n$ be defined
(\ref{burban: solu}) for $\gamma - \alpha > 0,$ $0 < q < 1$
$(\gamma - \alpha < 0, q > 1).$ The values of $\lambda_n, n \ge 0$
in (\ref{burban: fin}) are nonnegative if we assume
$\Bigl(\lambda_0 q^{-(\alpha \varkappa + \beta)} +
\frac{1}{q^{\gamma} - q^{\alpha}} + \frac{B}{q^{\gamma} +
q^{\alpha}}\Bigr) = 0$ and both values $\Bigl(\frac{1}{q^{\gamma}
- q^{\alpha}} + \frac{(-1)^n B}{q^{\gamma} + q^{\alpha}} \Bigr)$
are nonpositive. At $ B = \pm\frac{q^{\gamma} +
q^{\alpha}}{q^{\gamma} - q^{\alpha}}$ we obtain
\begin{equation}\label{burban: eveod}\lambda_n
= \frac{q^{\alpha(\varkappa_0 + n)+ \beta}}{q^{\alpha} -
q^{\gamma}}\Bigl(1 \pm (-1)^n\Bigr). \end{equation}

 The value $\lambda_n =
\frac{q^{\alpha(\varkappa_0 + n)+ \beta}}{q^{\alpha} -
q^{\gamma}}\Bigl(1 + (-1)^n)\Bigr)$  implies (\ref{burban:
findim1}) for $n_1 = - 1,$ $n_2 = 1$ and $\lambda_0 =
\frac{2q^{\alpha\varkappa_0 + \beta}}{q^{\alpha} - q^{\gamma}}.$

The vector space of this representation spanned by the
two-dimensional vectors $$ \left(
\begin{array}{c}
\psi_{-1} \\ \psi_0
\end{array}\right)
$$ and due to (\ref{burban: rep}) the representation are defined
by $$ a = \left(
\begin{array}{cc}
0&\sqrt{\frac{2 q^{\alpha\varkappa_0 + \beta}}{q^{\alpha} -
q^{\gamma}}}\\ 0& 0
\end{array}\right),\qquad
a^+ = \left(
\begin{array}{cc}
0& 0\\ \sqrt{\frac{2 q^{\alpha\varkappa_0 + \beta}}{q^{\alpha} -
q^{\gamma}}}& 0
\end{array}\right), $$
\begin{equation}
N = \left(
\begin{array}{cc}\chi_0 - 1
& 0\\0& \chi_0
\end{array}\right),\qquad K = \frac{1}{2\nu}\frac{q^{\gamma}+
q^{\alpha}}{q^{\alpha} - q^{\gamma}}\left(
\begin{array}{cc}
1&0\\0& 1
\end{array}\right).
\end{equation}
These representations are distinguished by the arbitrary values
$\varkappa_0,$ and $B = \pm\frac{q^{\gamma} +
q^{\alpha}}{q^{\gamma}- q^{\alpha}},$ $\lambda_0 =
\frac{2q^{\alpha(\varkappa_0 ) + \beta}}{q^{\alpha} -
q^{\gamma}}.$

The value $\lambda_n = \frac{q^{\alpha(\varkappa_0 + n) +
\beta}}{q^{\alpha} - q^{\gamma}}\Bigl(1 - (-1)^n\Bigr)$ implies
(\ref{burban: findim1}) for $n_1 = 0,$ $n_2 = 2.$ Moreover
$\lambda_1 = \frac {2 q^{\alpha(\varkappa_0 + 1) +
\beta}}{q^{\alpha} - q^{\gamma}}.$

The vector space of this representation spanned by the
two-dimensional vectors $$ \left(
\begin{array}{c}
\psi_0 \\ \psi_{-1}
\end{array}\right)
$$ and due to (\ref{burban: rep}) the representations are given by
$$ a = \left(
\begin{array}{c
c}
0&0\\\sqrt{\frac{q^{\alpha(\varkappa_0+1) + \beta}}{q^ {\alpha} -
q^{\gamma}}}& 0
\end{array}\right),\qquad
a^+ = \left(
\begin{array}{cc}
 0& \sqrt{\frac{q^{\alpha(\varkappa_0+1) + \beta}}{q^
{\alpha} - q^{\gamma}}}\\0& 0
\end{array}\right),
$$
\begin{equation}
N = \left(
\begin{array}{cc}
\chi_0& 0\\0& \chi_0 + 1
\end{array}\right),\qquad K = \frac{1}{2\nu}\frac{q^{\gamma}+
q^{\alpha}}{q^{\alpha} - q^{\gamma}}\left(
\begin{array}{cc}
-1&0\\0& 1
\end{array}\right).\qquad
\end{equation}
These representations are defined by arbitrary values of
$\varkappa_0$, and  $\lambda_0 = 0,$ $B = - \frac {q^{\gamma} +
q^{\alpha}}{q^{\gamma}- q^{\alpha}}$.

4. {\it Unbounded representations.} They defined by (\ref{burban:
rep}) with $\lambda_n \ge 0$ for all $ n\in{\mathbb Z}.$ As
follows from (\ref{burban: fin}) they are realized if $0 < q < 1,
(\alpha - \gamma)
> 0 \quad ( q > 1, (\alpha - \gamma) < 0 )$ and
\begin{equation}\label{burban: inf1} \lambda_0 q^{-\alpha \varkappa_0 - \beta} +
\frac{1}{q^{\gamma} - q^{\alpha}} + \frac {B}{q^{\alpha} +
q^{\gamma}}\ge 0,\quad |B| < - \frac{q^{\gamma} +
q^{\alpha}}{q^{\gamma} - q^{\alpha}},
\end{equation} or
\begin{equation} \label{burban: inf2} \lambda_0 q^{-\alpha \varkappa_0 - \beta} +
\frac{1}{q^{\gamma} - q^{\alpha}} + \frac {B}{q^{\alpha} +
q^{\gamma}}> 0, \quad |B|= -\frac{q^{\gamma} +
q^{\alpha}}{q^{\gamma} - q^{\alpha}}. \end{equation} Under the
transformations $$ B\to B'= (-1)^n B,\quad \varkappa_0\to
\varkappa' = \varkappa + n, $$
\begin{equation}\lambda_0\to \lambda'_0 =
\lambda_0 q^{\gamma n} + q^{\alpha\varkappa_0 + \beta}\Bigl(\frac
{q^{\gamma n} - q^{\alpha n}} {q^{\gamma} - q^{\alpha}} + B \frac
{q^{\gamma n} - (-1)^n q^{\alpha n}}{q^{\gamma} +
q^{\alpha}}\Bigr)
\end{equation}
the conditions (\ref{burban: inf1}) and (\ref{burban: inf2}) are
preserved
\begin{equation}
\lambda'_n = \lambda_0 q^{2\gamma n} + q^{\alpha\varkappa_0 +
\beta}\Bigl(\frac {q^{2\gamma n} - q^{2\alpha n}} {q^{\gamma} -
q^{\alpha}} + B \frac {q^{2\gamma n} - q^{2\alpha n}} {q^{\gamma}
+ q^{\alpha}}\Bigr),
\end{equation}
\begin{equation}
\lambda'_0 q^{-\alpha \varkappa'_0 - \beta} + \frac{1}{q^{\gamma}
- q^{\alpha}} + \frac {B'}{q^{\alpha} + q^{\gamma}} =
q^{(\gamma-\alpha)n}\Bigl(\lambda_0 q^{-\alpha \varkappa_0 -
\beta} + \frac{1}{q^{\gamma} - q^{\alpha}} + \frac {B}{q^{\alpha}
+ q^{\gamma}}\Bigr)
\end{equation}
and {\it unbounded representations} pass to themselves.


\noindent

{\bf 4. Spectrum of "quantized coordinate" $Q$}

\medskip

According to previous section  if we put $\omega = 1, \varkappa_0
= 0,$ in (\ref{burban: so1}) we obtain the Fock representation of
the relations (\ref{burban: first})
\[N|n\rangle = n|n\rangle,\quad K|n\rangle = (-1)^n|n\rangle, $$ $$
a^+|2n\rangle = q^{\beta/2} \Bigl(\frac{q^{\gamma (2n+1)} -
q^{\alpha (2n+1)}}{q^{\gamma} - q^{\alpha}} + 2\nu\frac{q^{\gamma(
2n+1)}+ q^{\alpha (2n+1)}}{q^{\gamma} + q^{\alpha
}}\Bigr)^{1/2}|2n+1\rangle,\]
\begin{equation} \label{burban: rep2} a^+|2n-1\rangle = q^{\beta/2} \Bigl(\frac{q^{2\gamma n} -
q^{2\alpha n}}{q^{\gamma} - q^{\alpha}} + 2\nu \frac{q^{2\gamma n}
- q^{2\alpha n}}{q^{\gamma} +
q^{\alpha}}\Bigr)^{1/2}|2n\rangle.\end{equation}

The properties of the quantum harmonic oscillator are closely
related to Hermite polynomials. Different $q$-deformed oscillators
generate distinct $q$-deformed Hermite polynomials and their
generalization. Unlike to quantum harmonic oscillator, generalized
q-deformed oscillators frequently define the position and momentum
operators with the discrete spectrum and their eigenfunctions are
represented by means of a q-Hermite polynomials. For some
q-deformed oscillators this problem was studied in \cite{BK},
\cite{CK}, \cite{K}. In \cite{Burb1} the generalized q-deformed
oscillators connected with the discrete (generalize) q-Hermite I
and q-Hermite II polynomials have been build. In this Section we
show that corresponding oscillator algebras of this models are
embedded in the $(q;\alpha,\beta,\gamma:\nu)$-deformed oscillator
algebra.

 Let us consider the operator ("quantized coordinate") $Q
= a^+ + a,$ or
\begin{equation}\label{burban: coord} Q|n\rangle = r_n|n+1\rangle + r_{n-1}|n-1\rangle,
\quad r_n = f^{1/2}(n+1).
\end{equation}

The self-adjointness and spectral properties of this operator is
defined by polynomials $P_n(x)$ of first kind for the Jacobi
matrix $Q.$

Defining the generalized eigenfunction $Q |x\rangle = x|x\rangle
$, where $|x\rangle = \sum_{n = 0}^{\infty} P_n(x)|n\rangle,$ we
obtain the recurrence relation
\begin{equation}\label{burban: jac}
f^{1/2}(n) P_{n-1}(x) + f^{1/2}(n+1) P_{n+1}(x) = x P_n(x),
\end{equation}
where $ P_{-1}(x) = 0, P_{0} = 1.$

In this Section we consider of the oscillator (\ref{burban:
first}) for special choice of parameters
\begin{equation}
\alpha = 2 a,\quad \gamma = 2 a + c - 1 \quad \beta = 2 a + b,\nu
= 0.
\end{equation}
We obtain the relation \cite {Burb1}:
\begin{equation}
\label{burban: grelation1} a a^+ - q^{2a + c-1} a^+ a = q^{2 a (N
+ 1) + b},
\end{equation}
or equivalent
\begin{equation}
\label{burban: grelation2} a a^+ - q^{2a} a^+ a = q^{2 a (N + 1) +
b}q^{(c-1)N} \end{equation} which together with other relations of
(\ref{burban: first}) form the corresponding oscillator algebra.

The self-adjoints of the operator (\ref{burban: coord}) can be
established by the Theorem 1.3 Chapter VII in \cite {Ber}. For
this purpose it is sufficient to prove the divergence of the
series
\begin{equation}\label{burban: karl}
\sum_{n = 0}^{\infty} \frac{1}{r_n} = \infty.
\end{equation}
The operator $Q$ (\ref{burban: coord}) is symmetric one. If
\begin{equation}
r_n\ge 0,\quad r_{n-1}r_{n+1}\le r_n^2,\quad \sum_{n=0}^{\infty}
\frac{1}{r_n} < \infty.
\end{equation}
by the Theorem 1.5 Chapter VII in \cite {Ber} its closure ${\bar
Q}$ is not self-adjoint operator.

In our case the property $ r_n\ge 0,\quad r_{n-1}r_{n+1}\le r_n^2$
is satisfied automatically, therefore, the self-adjointness of the
operator $Q$ are reduced to the proving of the convergence of the
series (\ref{burban: karl}). We obtain the cases
\begin{equation}
\label{burban:cas1} q < 1,\quad  \begin {cases} a < 0,\quad {c-1}
> 0,& \text{divergent,}\\ a > 0,\quad c-1 > 0, &\text{
convergent,}\\ a + \frac{c-1}{2} < 0,\quad c-1 <
0,&\text{divergent,}\\ a + \frac{c-1}{2} > 0,\quad c-1< 0, &
\text{convergent.}
\end{cases}
\end{equation}
The equation (\ref{burban: jac}) for this case take the form $$ x
P_n{(x;q)} = \Bigl(\frac{1}{1-q')}\Bigr)^{1/2} q^{a n +
b/2}(1-q'^{n})^{1/2}P_{n-1}(x;q)$$
\begin{equation}
\label{burban:equl1} + \Bigl(\frac{1}{1-q'}\Bigr)^{1/2}q^{{a(n +
1)} + b/2}(1-q'^{(n + 1)})^{1/2}P_{n + 1}(x;q),
\end{equation}
where $q'=q^{c-1}.$ If we  change and rescale of the variables $y
= (1-q')^{1/2} x, $ $ P_n(x;q)= \psi_n ((1 - q')^{1/2}x;q)$ then
it follows  $$ x \psi_n(x;q)$$
\begin{equation}
\label{burban:equer} = q^{a (n + 1) + b/2}(1 -
q'^{n})^{1/2}\psi_{n+1}(x;q) + q^{a n + b}(1 -
q'^{n})^{1/2}\psi_{n-1}(x;q).
\end{equation}
Representing $\psi_n(x;q)$ as $$\psi_n(x;q) = \frac {q^{a
n^2/2}}{q^{(a + b) n/2}(q';q')_n^{1/2}}h_n(x;q)$$ we get from
(\ref{burban:equer}) the recurrence relation for the (generalized)
discrete q-Hermite I polynomials $h_n(x;q):$
\begin{equation}
\label{burban: discrhermit} x h_n(x;q)= h_{n+1}(x;q)+ q^{2a n + b}
(1-q'^{n})h_{n-1}(x;q).
\end{equation}
The solution of this equation is given by anzatz
\begin{equation}
\label{burban:finitaI} h_n(x;q)= \sum_{k = 0}^{[n/2]}
\frac{(q';q')_n}{\Bigl((a_n,c_n); (1,q'^d)\Bigr)_k}(-1)^k q^{(2a n
+ b)k} q'^{k(k-n)} x^{n-2k}
\end{equation}
where $a_n, c_n, d$ are unknown quantities.  The notation
\begin{equation}
\Bigl((a, c );(p,q)\Bigr)_k = \begin {cases} 1,& \text{if $ k = 0
;$}\\ (a - c)(a p - c q)\ldots(a p^{k-1}- c
q^{k-1}),&\text{otherwise }\\
\end{cases}
\end{equation} we use from \cite{JR}.
The substitution (\ref{burban:finitaI}) into (\ref{burban:
discrhermit}) leads to the equation
\begin{equation}
1-q'^{n+1 }- q^{2a n + b}(a_n-c_n q'^{d(k-1)}) q'^{-2(k-1)} = 1 -
q'^{n-2k+1}
\end{equation}
The solutions of this equation
\begin{equation}
a_n = q^{-2a n-b}q'^{n-1}, c_n = q^{-2a n-b}q'^{n+1}, d+2
\end{equation}
give the solutions of (\ref{burban: discrhermit})
\begin{equation} h_n(x;q)= \sum_{k = 0}^{[n/2]}
\frac{(q';q')_n}{(q'^2 ; q'^2)_k (q';q')_{n-2k}} (-1)^k q^{(2a n +
b)k} q'^{k(k-n)} x^{n-2k}.
\end{equation}
They can be represented in terms of the basic hypergeometric
function as
\begin{equation}
h_n(x;q)= x^n{}_2\phi_0\Bigl( \begin{array}{cc|c}
q'^{-n},&q'^{-n+1}&\\ {}&-&\end{array}q'^2;\frac{q^{2a n + b}
q'^n}{x^2} \Bigr).
\end{equation}
 As has been shown in \cite {Burb1} these solutions at $a = \frac{1}{2}, b = - 1, c =
2$ is reduced to discrete q-Hermite I polynomials and the
relations (\ref{burban: grelation1}) (or (\ref{burban:
grelation2})) to equation of the corresponding $q$-deformed
oscillator.

 The solutions $P_n(x;q)$ of the equations
(\ref{burban:equl1}) with the initial conditions $P_{-1}(x;q) =
0,\quad P_0(x;q) = 1$ can be written as polynomials of degree $n$
in $x:$
\begin{equation} \label{burban:resI} P_n(x;q) =
\frac{q^{- a n^2/2}}{q^{\frac{a + b}{2} n}(q';q')_n^{1/2}}h_n
(\sqrt{1-q'}x;q).
\end{equation}
\medskip
Now we restrict ourselves by the condition $a = (c-1)/2$ in
(\ref{burban:resI}). Then \begin{equation} P_n(x;q)= \frac{q^{-a
n(n-1)/2}}{(q';q')_n ^{1/2}}h_n^0 (q^{-(2a + b)/2}
\sqrt{1-q'}x;q').
\end{equation}
These polynomials are orthogonal with respect to the discrete
measure $$ d\omega(x) = \frac {q^{-(2a + b)}\sqrt{1-q'}
}{2}(q';q'^2)_{\infty}\delta (x - \frac {q'^0}{ q^{-(2 a + b
)}\sqrt{1-q'}})d x $$ $$+ \sum_{k > 0} \frac{q^{-(2 a +
b)/2}\sqrt{1-q'}|x|}{2}\frac{(q^{-(2 a + b)}(q'^2(1-q')x^2,q';
q^2)_{\infty}}{(q';q')_{\infty}} \,\delta (x - \frac {q'^k}{q^{-(2
a + b)/2}\sqrt{1-q'}})d x $$
\begin{equation}
+ \sum_{k > 0} \frac{q^{-(2a +
b)/2}\sqrt{1-q'}|x|}{2}\frac{(q^{-(2 a + b)}(q'^2(1-q')x^2,q';
q^2)_{\infty}}{(q';q')_{\infty}} \,\delta ( x + \frac {q'^k}{q^{-(
2 a + b)/2}\sqrt{1-q'}})d x.
\end{equation}
The orthogonality relation has the form $$
\frac{\delta_{mn}}{(q';q')_n} =
\frac{1}{2}\frac{(q';q')_{\infty}}{(q'^2;q'^2)_{\infty}}
P_m(1;q)P_n(1;q)  $$
\begin{equation}
+ \sum_{k > 0}\{P_m (q'^k;q)P_n(q'^k;q) + P_m (- q'^k;q)P_n(-
q'^k;q)\}\frac{q'^k}{2}\frac{(q'^{2k+2},q';q'^2)_{\infty}(q';q'^2)_{\infty}}
{(q'^2;q'^2)_{\infty}}.\end {equation} It follows that spectrum of
the position operator $Q$ is
\begin{equation} Sp\,Q =
\Bigl\{ \frac{\pm q^{(2a + b)/2}}{\sqrt{1-q'}},\quad\frac{\pm
q^{(2a + b)/2}q'}{\sqrt{1-q'}},\ldots, \frac{\pm q^{(2 a +
b)/2}q'^k}{\sqrt{1-q'}},\ldots; k\ge 0 \Bigr\}.
\end{equation}


\noindent

{\bf 4. Hamiltonian and energy spectrum of
$(q;\alpha,\beta,\gamma;\nu)$-deformed oscillator}

\medskip

{\it (g) Energy spectrum of free
$(q;\alpha,\beta,\gamma;\nu)$-deformed oscillator.} In general, a
Hamiltonian $H(a^+,a,N)$ of an oscillator is a function of the
operators $a^+, a, N.$ As an example, we consider the free
standard Hamiltonian
\begin{equation}\label{burban: ham} H =
\frac{\hbar\omega_0}{2}\{a^+, a\} = \frac{\hbar\omega_0}{2}(aa^+ +
a^+a)\end{equation} of the $(q;\alpha,\beta,\gamma;\nu)$-deformed
oscillator.

Due to the structure function of deformation the free Hamiltonian
$H$ of the $(q;\alpha,\beta,\gamma;\nu)$-deformed oscillator in
the Fock representation (\ref{burban: rep2}) can be written only
as function of number operator $N$
\[ H = \frac{\hbar\omega_0}{2}q^{\beta}\Bigl\{
\Bigl(\frac{q^{\gamma N} - q^{\alpha N}}{q^{\gamma} - q^{\alpha}}
+ 2 \nu \frac{q^{\gamma N} - (-1)^N q^{\alpha N}}{q^{\gamma} +
q^{\alpha}}\Bigr) +
\]
\begin{equation} \Bigl(\frac{q^{\gamma (N+1)} -
q^{\alpha (N+1)}}{q^{\gamma} -
q^{\alpha}} + 2 \nu \frac{q^{\gamma (N+1)}-(-1)^{(N+1)} q^{\alpha
(N+1)}}{q^{\gamma} + q^{\alpha}}\Bigr)\Bigr\}.
\end{equation}

For a study of the spectrum and the nonlinear frequency of this
oscillator, it is convenient to use new parametrization
\begin{equation}\label{burban: newpar}
q = e^{\tau},\quad \alpha = \rho-\mu,\quad \gamma = \rho
+\mu.\end{equation}
We have
\[ E_n = \frac{\hbar\omega_0}{2}e^{\tau(\beta + \rho n)}\times \]
\[ \Bigl\{\frac{\sinh(\tau\mu(n+1))}{\sinh(\tau\mu)} +
e^{-\tau\rho}\frac{\sinh(\tau\mu n)}{\sinh(\tau \mu)} +
2\nu\frac{1 -
(-1)^n}{2}\Bigl(\frac{\sinh(\tau\mu(n+1))}{\cosh(\tau \mu)} +
e^{-\tau \rho} \frac{\cosh(\tau \mu n)}{\cosh(\tau \mu)}\Bigr) +
\]
\begin{equation}\label{burban: hamo}
2\nu\frac{1 + (-1)^n}{2}\Bigl(\frac{\cosh(\tau \mu (n +
1))}{\cosh( \tau \mu)} + e^{-\tau\rho}\frac{\sinh(\tau\mu
n)}{\cosh(\tau \mu)}\Bigr) \Bigr\},
\end{equation}
or
\[ E_n = \frac{\hbar\omega_0}{2}e^{\tau(\beta - \rho)}
\Bigl\{ \frac{1 + e^{\tau(\rho + \mu)}}{2}\frac{e^{\tau(\rho +
\mu)n}}{\sinh(\tau\mu)} - \frac{1 + e^{\tau(\rho -
\mu)}}{2}\frac{e^{\tau(\rho - \mu)n}}{\sinh(\tau\mu)}
\]
\[ + 2 \nu
\frac{1 - (-1)^n}{2}\Bigl( \frac{1 + e^{\tau(\rho +
\mu)}}{2}\frac{e^{\tau(\rho + \mu)n}}{\cosh(\tau \mu)} + \frac{1 -
e^{\tau(\rho - \mu)}}{2}\frac{e^{\tau(\rho - \mu)n}}{\cosh(\tau
\mu)}\Bigr)+\]
\begin{equation}\label{burban: hami}2\nu \frac{1 + (-1)^n}{2}\Bigl(\frac{1 + e^{\tau(\rho +
\mu)}}{2}\frac{e^{\tau(\rho + \mu)n}}{\cosh(\tau\mu)} - \frac{1 -
e^{\tau(\rho - \mu)}}{2}\frac{e^{\tau(\rho -
\mu)n}}{\cosh(\tau\mu)}\Bigr)\Bigr\}.
\end{equation}
It follows that the number of the energy levels is infinite, and
the eigenvalues $E_n$ for $\tau \ne 0, n\to\infty$ depend on the
parameters $\alpha, \gamma$ via the exponential factor
$e^{\tau(\rho - |\mu|)}.$

It is convenient to consider the eigenvalue $E_n$ of $H$ for $n$
even and odd separately:
\[
E_n = \frac{\hbar\omega_0}{2}e^{\tau(\beta - \rho)}\Bigl\{\frac{1
+ e^{\tau(\rho + \mu)}}{2}\frac{e^{\tau(\rho +
\mu)n}}{\sinh(\tau\mu)} - \frac{1+e^{\tau(\rho -
\mu)}}{2}\frac{e^{\tau(\rho - \mu)n}}{\sinh(\tau\mu)} + \]
\begin{equation}
\label{burban: odd} 2\nu\Bigl(\frac{1 + e^{\tau(\rho +
\mu)}}{2}\frac{e^{\tau(\rho + \mu)n}}{\cosh(\tau\mu)}+\frac{1 -
e^{\tau(\rho - \mu)}}{2}\frac{e^{\tau(\rho -
\mu)n}}{\cosh(\tau\mu)}\Bigr)\Bigr\}
\end{equation}
for $n$ odd, and
\[ E_n = \frac{\hbar \omega_0}{2}e^{\tau(\beta - \rho)}\Bigl\{\frac{1
+ e^{\tau(\rho + \mu)}}{2}\frac{e^{\tau(\rho +
\mu)n}}{\sinh(\tau\mu)} - \frac{1 + e^{\tau(\rho -
\mu)}}{2}\frac{e^{\tau(\rho - \mu)n}}{\sinh(\tau\mu)}+ \]
\begin{equation}\label{burban: even}
2\nu\Bigl(\frac{1 + e^{\tau(\rho + \mu)}}{2}\frac{e^{\tau(\rho +
\mu)n}}{\cosh(\tau\mu)} - \frac{1 - e^{\tau(\rho -
\mu)}}{2}\frac{e^{\tau(\rho - \mu)n}}{\cosh(\tau\mu)}\Bigr)\Bigr\}
\end{equation}
for $n$ even. It follows the spectrum of this oscillator is not
equidistant and the spacing is equal

\begin{equation}
E_{2n+1} - E_{2n} = \frac{\hbar
\omega_0}{2}\frac{\sinh2\mu\tau}{\sinh(\tau\mu)} e^{\tau(\beta
+\rho)} e^{2\gamma\tau n}.
\end{equation}

{\it (h) Asymptotic behaviour of energy spectrum of Hamiltonian
$H$.}

 According to the analysis given above, we have
to give the estimation of the spectrum of Hamiltonian (\ref
{burban: hami}) for the parameter $-1 < 2\nu ,$ if $\tau(\gamma -
\alpha) > 0$ and $-1 < 2 \nu < \coth(\tau(\gamma - \alpha)/2)$ if
$\tau(\gamma - \alpha) < 0.$
 Three cases must be distinguished:
$\tau(\rho + |\mu|)>0,\tau(\rho + |\mu|)< 0,\tau(\rho + |\mu|) =
0.$

In special case $\mu = 0,$ $(\alpha = \gamma = \rho) $ we have
\begin{equation}
E_n =\frac{\hbar\omega_0}{2}e^{\tau(\beta + \rho
n)}\Bigl[(n+\nu)(1 + e^{-\tau\rho})+(-1)^n\nu(1 - e^{-\tau\rho}) +
1 \Bigr].
\end{equation}

If $(\tau(\rho + |\mu|)
> 0$ and $ -1 < 2\nu$, then from (\ref{burban: hami}) it implies that
the energy grows to infinity with increasing $n.$

If $\tau(\rho + |\mu|) = 0,$ then in the case ($\rho = \mu $) we
have
\begin{equation}
E_n = \frac{\hbar \omega_0}{2}e^{\tau \beta}e^{\beta -
\rho}\Bigl\{ \frac{e^{2\tau\rho n} -1}{e^{\tau\mu}-e^{-\tau\mu}} +
\frac{e^{2\tau\rho (n+1)} -1}{e^{\tau\mu} - e^{-\tau\mu}} + 2 \nu
\frac{e^{2\rho\tau n} - e^{2\rho\tau (n+1)}}{e^{\tau\mu}-
e^{-\tau\mu}}\Bigr\}.
\end{equation}
The same holds for $\rho +\mu = 0.$

It follows that $E_n$ monotonically increase to upper bounds that
depends on parameters

$$E_{\max} = \frac{\hbar\omega_0}{2}\frac{e^{\tau (\beta -
\rho)}}{\sinh\tau|\mu|}.$$

If $\tau(\rho + |\mu|) < 0$ and $ -1 < 2\nu$ then the energies
(\ref{burban: odd}) and (\ref{burban: even}) at first increase and
then go to zero as $n\to\infty.$ In the particular case $\mu = 0$
in (\ref{burban: odd}) and (\ref{burban: even}) we have

\[E_n = \frac{\hbar \omega_0}{2}e^{\tau(\beta + \rho n)}\Bigl\{n(1+
e^{- \tau\rho}) + 2 \nu + 1\Bigr\},\] for $n$ odd and
\[ E_n = \frac{\hbar \omega_0}{2}e^{\tau(\beta + \rho n)}\Bigl\{n (1 + e^{-\tau\rho}) +
2 \nu e^{- \tau\rho} + 1 \Bigr\}\] for $n$ even.

\subsection*{Acknowledgements}

This research was partially supported by Grant 14.01/016 of the
State Foundation of Fundamental Research of Ukraine.

\end {document}